\documentclass[aps,prb,twocolumn,superscriptaddress]{revtex4-1}
\usepackage{graphicx}
\usepackage{color}
\usepackage{tabularx}
\usepackage{multirow}
\usepackage{amsfonts}
\usepackage{amssymb}
\usepackage[]{amsmath}
\usepackage[dvipsnames]{xcolor}	
\usepackage[]{natbib}
\usepackage[bookmarks=true,															
bookmarksnumbered=true,										
colorlinks=true,
urlcolor=black,
linkcolor=blue,
citecolor=blue]{hyperref}		    	
\newcommand*{\sect}{Sec.\,}

\newcommand*{\fig}{Fig.\,}
\newcommand*{\eq}[1]{Eq.\,(#1)}
\newcommand*{\tab}{Table\,}

\newcommand*{\ie}{\emph{i.e.}}
\renewcommand{\vec}[1]{\boldsymbol{#1}}

\newcommand*{\eg}{e.g.\,}

\graphicspath{ {Figures/} }

\newcommand{\ceptal}{CePtAl$_{3}$}

\newcommand{\neel}{N\'{e}el}

\newcommand{\JCNSMLZ}{J\"{u}lich Centre for Neutron Science (JCNS) at Heinz Maier-Leibnitz Zentrum (MLZ), Forschungszentrum J\"{u}lich GmbH, Lichtenbergstr. 1, D-85747 Garching}
\newcommand{\TUM}{Physik-Department, Technische Universit\"{a}t M\"{u}nchen (TUM), D-85748 Garching, Germany}
\newcommand{\CUni}{Charles University, Faculty of Mathematics and Physics, Department of Condensed Matter Physics, Ke Karlovu 5, 121 16, Praha, Czech Republic}
\newcommand{\MLZTUM}{MLZ, TUM, D-85748 Garching, Germany}
\newcommand{\PSI}{Paul Scherrer Institut, Villigen, Switzerland}
\newcommand{\UGre}{University Grenoble Alpes, CEA, IRIG, MEM, MDN, 38000 Grenoble, France}
\newcommand{\JCNSILL}{JCNS at Institute Laue-Langevin, Forschungszentrum J\"{u}lich GmbH, 38042 Grenoble, France}
\newcommand{\ISIS}{ISIS Neutron and Muon Source, STFC Rutherford Appleton Laboratory, Harwell Campus, Didcot, Oxon OX11 0QX, United Kingdom}
\newcommand{\Aachen}{Institut f\"{u}r Kristallographie at MLZ, RWTH Aachen, Lichtenbergstrasse 1, D-85747 Garching}
\newcommand{\ZQE}{Center for QuantumEngineering (ZQE), TUM, 85748 Garching, Germany}
\newcommand{\MCQST}{Munich Center for Quantum Science and Technology (MCQST), TUM, 85748 Garching, Germany}

\begin{document}

\title{Incommensurate antiferromagnetic order in CePtAl$_3$}

\author{M.~Stekiel}
\email{michal.stekiel@frm2.tum.de}
\email{m.stekiel@fz-juelich.de}
\affiliation{\JCNSMLZ}
\affiliation{\TUM}

\author{P.~\v{C}erm\'{a}k}
\affiliation{\JCNSMLZ}
\affiliation{\CUni}

\author{C.~Franz}
\affiliation{\JCNSMLZ}
\affiliation{\MLZTUM}

\author{W.~Simeth}
\affiliation{\PSI}

\author{S.~Weber}
\affiliation{\TUM}

\author{E.~Ressouche}
\affiliation{\UGre}

\author{W.~Schmidt}
\affiliation{\JCNSILL}

\author{K.~Nemkovski}
\affiliation{\JCNSMLZ}
\affiliation{\ISIS}

\author{H.~Deng}
\affiliation{\JCNSMLZ}
\affiliation{\Aachen}

\author{A.~Bauer}
\affiliation{\TUM}
\affiliation{\ZQE}

\author{C.~Pfleiderer}
\affiliation{\TUM}
\affiliation{\ZQE}
\affiliation{\MCQST}

\author{A.~Schneidewind}
\affiliation{\JCNSMLZ}


\date{\today}

\begin{abstract}
We report a neutron diffraction study of single-crystal CePtAl$_3$ complemented by measurements of the specific heat under applied magnetic field. Below $T_\mathrm{N}\approx3$\,K CePtAl$_3$ develops incommensurate antiferromagnetic order with a single modulation vector $\vec{k}$=$(0.676 \, 0 \, 0)$. Residual magnetic scattering intensity above $T_\mathrm{N}$ and a broadening of the specific heat anomaly at $T_\mathrm{N}$ may be consistently described in terms of a Gaussian distribution of transition temperatures with a standard deviation $\sigma \approx 0.5$\,K. The distribution of $T_\mathrm{N}$ may be attributed to the observation of occupational and positional disorder between the Pt and Al sites consistent with structural information inferred from neutron diffraction. Measurements under magnetic field reveal changes of the magnetic domain populations when the field is applied along the $[0\,1\,0]$ direction consistent with a transition from cycloidal to amplitude-modulated magnetic order around 2.5\,T.
\end{abstract}

\maketitle

\section{Introduction}

Cerium based intermetallics comprise a wide variety of physical phenomena owing to the coupling between electronic and magnetic degrees of freedom and their relationship to the crystal structure and its symmetry. Perhaps most extensively studied are materials with the ThCr$_2$Si$_2$ structure, featuring heavy fermion superconductivity, quantum phase transitions, as well as large magneto-elastic coupling phenomena \cite{Shatruk-JSSC-2019, Pfleiderer-RevModPhys-2009}. Born out of this long history, a structurally related materials class that has attracted increasing interest in recent years comprises Ce$TX_3$ systems, where $T$ is a transition metal and $X=$ Al, Ga, Si, or Ge. For instance, in CeCuAl$_3$ and CeAuAl$_3$ anomalous thermal transport properties and inelastic neutron and x-ray scattering experiments have revealed magnetoelastic hybrid excitations even for nominally weak magnetoelastic coupling \cite{Paschen-EPJ-1998, Adroja-PRB-2015, Cermak-PNAS-2019}. Moreover, antiferromagnetic (AFM) order with low transition temperatures have been reported, \eg, for $T=$ Cu and Au, where CeCuAl$_3$ exhibits an amplitude modulated, collinear, AFM structure, with a modulation vector $\vec{k}=(\frac{2}{5}\,\frac{3}{5}\,0)$ below $T_\mathrm{N}$\,=\,2.7\,K \cite{Klicpera-PRB-2015}, while CeAuAl$_3$ exhibits a helical AFM structure with $\vec{k}=(0\,0\,0.52)$ below $T_\mathrm{N}$\,=\,1.2\,K \cite{Adroja-PRB-2015}.

A key aspect of the $I4mm$ space group, in which most members of the Ce$TX_3$ class crystallize, is the presence of a four-fold symmetry axis along the $[0\,0\,1]$ direction. For $\vec{k}$ vectors in the basal plane as in CeCuAl$_3$ this permits, in principle, the formation of multi-$k$ magnetic structures. This possibility was considered unlikely for CeCuAl$_3$ \cite{Klicpera-PRB-2015}; however, multi-$k$ structures were observed in various other cerium intermetallics \cite{Schweizer-JOP-2008, Puphal-PRL-2020, Donni-JOP-1993}. Thus, to unambiguously determine the magnetic state, the number of modulation vectors must be determined, and measurements under symmetry-breaking conditions must be performed to distinguish a potential multi-$k$ structure from the formation of magnetic domains. On this note, the putative presence of multi-$k$ order in the Ce$T$Al$_3$ family would identify another exciting facet of correlated electron systems. Further, the low \neel\ temperatures in these systems suggest a possible vicinity to a quantum phase transition. This raises the question if other members of the Ce$T$Al$_3$ family exist, that are either nonmagnetic in the zero temperature limit or even coincidentally located at a quantum critical point. 

In this paper we report the observation of long-range AFM order in CePtAl$_3$. This system crystallizes in the BaNiSn$_3$-type structure, with a non-centrosymmetric, tetragonal $I4mm$ space group\cite{Franz-JAC-2016}. Previous reports on the possible existence of long-range magnetic order in CePtAl$_3$ were inconclusive \cite{Mock-JLTP-1999, Franz-JAC-2016}. The electrical resistivity, magnetization, and specific heat of polycrystalline samples were found to be characteristic of spin-glass behavior $\le$0.8\,K \cite{Mock-JLTP-1999}. However, the authors of that study noted \cite{Mock-JLTP-1999} that their CePtAl$_3$ sample showed a high degree of crystalline disorder, speculating on an intrinsic lack of long-range magnetic order. Likewise, based on the electrical resistivity, \citet{Franz-JAC-2016} suggested that CePtAl$_3$ does not exhibit magnetic order down to 0.1\,K. Moreover, the magnetization of single-crystal CePtAl$_3$ shows an almost linear magnetic field dependence up to 9\,T \cite{Franz-PhD}, without signs of saturation. The magnetic moments at 9\,T and 2\,K are 1 and 0.3\,$\mu_\mathrm{B}$ for field applied along the $a$ and $c$-axis, respectively, which is far from the value expected for an isolated Ce$^{3+}$ atom of 2.14\,$\mu_\mathrm{B}$. This suggests the presence of crystal field effects that influence the magnetism in CePtAl$_3$ and establishes an easy magnetization plane $(0\,0\,1)$ and a hard magnetic axis along the $[0\,0\,1]$ direction. Taken together, these data suggested a possible vicinity to a quantum phase transition, motivating the detailed microscopic study reported in this paper.


\section{Experimental methods}
\label{sect:methods}

Centimeter-sized single crystals of CePtAl$_3$ were grown from polycrystalline rods using an optical floating-zone furnace, described in more detail in Refs.\,\onlinecite{Franz-PhD, Neubauer-RSI-2011, Bauer-RSI-2016}. Both the stability of the floating-zone growth, which was well-behaved, and the metallurgical appearance of the ingot reject the possibility of macroscopic phase segregation. Rather, behavior characteristic of homogeneous sample was observed. The crystals measured in this study were prepared from the same batch studied by \citet{Franz-JAC-2016}. Samples were cut from the single-crystalline rod as required for the different experimental techniques. Demagnetization effects were negligible for all samples studied.
%
The specific heat was measured in a Quantum Design physical properties measurement system at temperatures down to 2\,K under applied magnetic fields up to 14\,T. Constant field scans as a function of temperature were performed on a 1$\times$1$\times$1\,mm$^3$ single-crystal cube of CePtAl$_3$ using a large heat-pulse method. Constant temperature scans as a function of an applied magnetic field were carried out on a 2$\times$1$\times$6\,mm$^3$ single crystal using a small heat-pulse method. In all specific heat measurements, the magnetic field was applied along the $[1\,0\,0]$ direction.
%
Neutron diffraction was carried out at several different beamlines using a large cylinder-shaped crystal with a length and diameter of 12 and 6\,mm, respectively (volume=340\,mm$^3$).

Initial single-crystal neutron diffraction measurements were performed on the triple-axes-spectrometer PANDA \cite{PANDA-LSF} at the MLZ. The crystal was mounted in a $^{3}$He cryostat and cooled to 0.5\,K. The intensities of selected reflections were measured as a function of temperature to investigate the magnetic transition. The modulation vector of the magnetic order was determined from measurements at the Diffuse Neutron Scattering spectrometer (DNS) \cite{DNS-LSF, Schweika-PhysicaB-2001} at MLZ using polarization analysis. The incoming wavelength of 4.2\,\AA\ and the one-dimensional detector bank at the DNS allowed us to simultaneously collect data for a range of momentum transfers 0.25$<$$Q$$<$2.7\,\AA$^{-1}$. The sample was rotated by 1$^{\circ}$ steps in a 140$^{\circ}$ range to map out the neutron scattering intensity on a plane in reciprocal space. Here, $xyz$-polarization analysis was employed with the neutron polarization set parallel to the momentum transfer, which allowed us to distinguish between magnetic and nuclear scattering contributions on the spin-flip (SF) and non spin-flip (NSF) channels, respectively \cite{Schweika-JPhysConf-2010, Boothroyd-2020}. The crystal was cooled to 0.45\,K using a $^{3}$He cryostat. Data were analyzed and reduced with the \textsc{mantid} package \cite{Mantid} allowing us to project the measured intensities on selected crystal planes.

Further single-crystal neutron diffraction measurements were performed on POLI \cite{POLI-LSF} at MLZ and on D23 \cite{D23-LSF} at the ILL. Both diffractometers operate a sample stage with an $\omega$ rotation axis and a lifting detector stage allowing us to measure out-of-plane scattering intensity. Measurements at POLI aimed at an accurate determination of the crystal structure.

To minimize extinction effects, a small half-disc-shaped crystal with dimensions 6 $\times$ 1\,mm (diameter $\times$ thickness) and a volume of 14\,mm$^3$ was used at an incoming wavelength of 3.355\,\AA. Data were first recorded at ambient conditions. Following this, the sample was mounted in the $^{3}$He cryostat and cooled to 0.45\,K. Measurements at D23 aimed at the determination of the magnetic structure and were performed with a 6\,T vertical field magnet at 1.5\,K at an incoming wavelength of 1.275\,\AA. The same crystal that was used at the DNS and PANDA ($V$=340\,mm$^3$) was mounted in two configurations, first with the $[0\,1\,0]$ direction vertical and second with the $[0\,0\,1]$ direction vertical.

The experimental protocol for measurements under magnetic field at D23 was as follows. The sample was cooled to 1.5\,K in zero field, and the magnetic field was applied along the $[0\,1\,0]$ direction, \ie along the $b$ axis. The field was ramped from $\mu_0 H=0$ to 5.5\,T and then down to $\mu_0 H=0$ in steps. At each field, a series of $Q$ and $\omega$-scans was measured to determine the position and intensity of selected reflections. Following this, the sample was heated to ambient temperature, reoriented, and cooled down to 1.5\,K again, so that the magnetic field was applied along the $[0\,0\,1]$ direction, \ie, along the $c$ axis. For this orientation we measured magnetic and nuclear reflections up to 4\,T under increasing magnetic field. Integration of the reflections from the diffraction measurements was done with the \textsc{davinci} program \cite{Davinci}. For the determination of magnetic structures, refinements were carried out using \textsc{jana2006} \cite{Jana-2014} with the symmetry analysis input from \textsc{isodistort} \cite{Isodistort}.


\section{Results and Interpretation}

The presentation of our experimental results is organized as follows. The specific heat data are presented in \sect\ref{sect:cp}. Following this, the results of four independent neutron diffraction measurements are reported. Measurements focusing on the accurate determination of the crystal structure are presented in \sect\ref{sect:crystal}. The results of the search for the magnetic modulation vector and the characterization of the magnetic order parameter are presented in \sect\ref{sect:magnetic-transition}. Finally, determination of the AFM order and its behavior under applied magnetic fields are reported in \sect\ref{sect:magnetism}.


\subsection{Specific heat}
\label{sect:cp}

\begin{figure}[t]
	\begin{center}
		\includegraphics[width=0.48\textwidth]{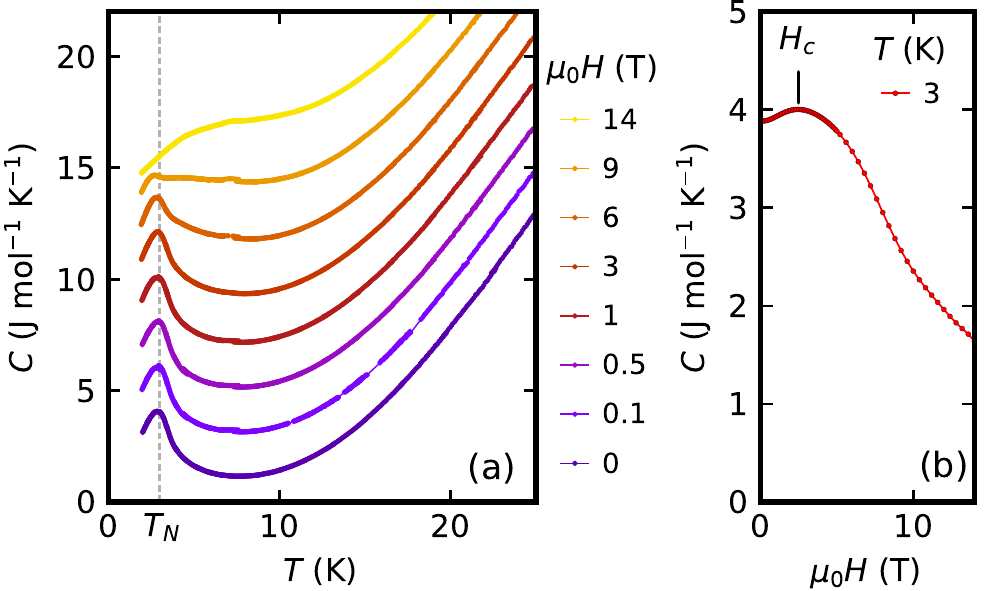}
	\end{center}
	\caption{Specific heat of CePtAl$_3$. (a) Specific heat as a function of temperature at various magnetic fields up to 14\,T. Data are shifted vertically by 2 J/(mol\,K) for clarity. The dashed line marks characteristic temperature $T_\mathrm{N}$ at which the field sweep was measured as shown in (b). In (a), the maximum in specific heat at $T_\mathrm{N}$=3\,K marks the transition to the magnetically ordered state, consistent with neutron scattering. (b) Specific heat as a function of magnetic field at $T$=3\,K. In (b), the local maximum of the 3~K field sweep is marked with $H_\mathrm{c}$ at 2.5~T and coincides with the magnetic transition observed with neutron scattering.}
	\label{fig:cp}
\end{figure}

The specific heat of CePtAl$_3$ was measured as a function of temperature at various applied magnetic fields up to 14\,T, as shown in \fig\ref{fig:cp}, where we focused on the low-temperature part in the regime of the magnetic transition. At zero magnetic field, the specific heat shows a broad maximum centered at $T_\textrm{N}$=3\,K with a full-width-at-half-maximum (FWHM) of $\approx$1.5\,K. Under increasing applied field, there is no visible change in the specific heat curve up to 3\,T. At 6 and 9\,T, the amplitude of the maximum is decreased and slightly shifted toward lower temperatures. The maximum has vanished at 14\,T.

The field dependence was explored further in measurements as a function of magnetic field at fixed temperature, as shown in \fig\ref{fig:cp}(b). At 3~K, the specific heat exhibits a maximum at 2.5~T marked as $H_\mathrm{c}$ and a monotonic decrease under increasing field above $H_\mathrm{c}$.

As shown below, neutron diffraction allows us to associate the broad maxima in the specific heat at 3\,K and zero field with the transition to AFM order, while the local maximum at $\mu_0 H_\mathrm{c}=$\,2.5\,T coincides with a field-driven magnetic transition.

\subsection{Crystal structure}
\label{sect:crystal}

\begin{figure}[t]
	\begin{center}
		\includegraphics[width=0.5\textwidth]{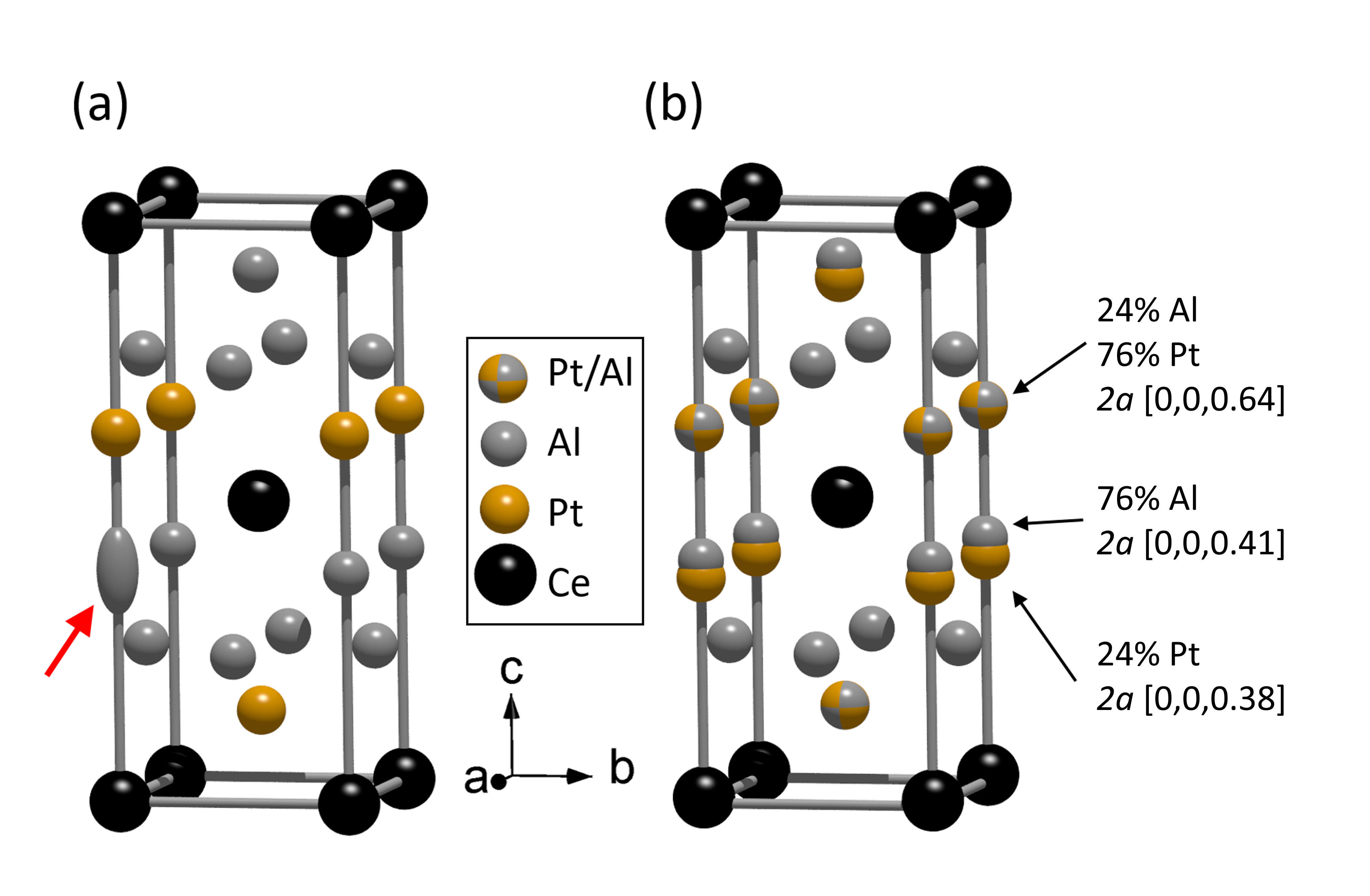}
	\end{center}
	\caption{Crystal structure of CePtAl$_3$. (a) Archetypal BaNiSn$_3$ structure. Red arrow in (a) points at the Al atom on the 2$a$ site which exhibits strong elongation of displacement ellipsoid and is modeled by splitting the equivalent sites (b) BaNiSn$_3$ structure modified with occupational and positional disorder on the Al 2$a$ and Pt 2$a$ sites. Occupations of the disordered sites and their positions are indicated by the labels.}
	\label{fig:crystal-structure}
\end{figure}

Single-crystal neutron diffraction at ambient conditions confirmed that \ceptal\ crystallizes in the BaNiSn$_3$ structure with a non-centrosymmetric $I4mm$ space group, as previously reported by \citet{Franz-JAC-2016} and shown in \fig\ref{fig:crystal-structure}. Lattice parameters and atomic positions are summarized in \tab\ref{tab:crystal-structure}. A large number of high-$Q$ reflections allowed for a reliable refinement of the displacement parameters.

The refinement for the BaNiSn$_3$ structure showed negative displacement parameters on the Al 2$a$ site and a rather high refinement quality factor $R_\mathrm{obs}$=12.24. The refinement was found to be improved, $R_\mathrm{obs}$=5.67, when assuming antisite disorder between the Al and Pt 2$a$ sites. Further, the displacement ellipsoid of the Al 2$a$ site was found to be strongly elongated along the $c$ axis, while other sites were found to show a fairly isotropic displacement, as illustrated in \fig\ref{fig:crystal-structure}(a). This was taken as an indication of positional disorder. Splitting, in turn, the Al 2$a$ site into separate sites occupied by Pt and Al atoms gave a final figure of merit $R_\mathrm{obs}$=4.98, where the occupancies were constrained to retain the stoichiometric composition of CePtAl$_3$. Refinements without such a constraint did not improve the quality of the fit. The final model is summarized in \tab\ref{tab:crystal-structure} and illustrated in \fig\ref{fig:crystal-structure}(b).

\begin{table}[t]
	\centering
	\caption{Details of the CePtAl$_3$ crystal structure refinement based on neutron diffraction at ambient temperature. Space group $I4mm$, $a$=$b$=4.326\,\AA,  $c$=10.655\,\AA. The number of measured:independent:observed reflections is 343:175:156, where $R_\mathrm{int}$=8.21, and observed reflections are with $I>3\sigma(I)$. The refinement was performed on $F^2$ with $R_\mathrm{obs}$=4.98 and $wR_\mathrm{obs}$=11.11. The columns contain the symbol of the Wyckoff site, atoms occupying the site, occupancy, $z$ coordinate of the Wyckoff site (2$a$: [0, 0, $z$] and 4$b$:[0, 0.5, $z$]), and associated, refined displacement parameters, respectively.}
	\label{tab:crystal-structure}
	\begin{tabular}{l c c c c}
		\hline\hline
		Site & Atom & Occupancy & $z$ & U (\AA$^2$) \\
		\hline
		\multirow{2}{*}{2$a$} & \multirow{2}{*}{Ce} & \multirow{2}{*}{1} & \multirow{2}{*}{0 (fixed)} & U$_{11}$=0.0168 (17) \\
		&  &  &  & U$_{33}$=0.0158 (20) \\
		\hline
		\multirow{2}{*}{2$a$} & Pt & 0.76 & \multirow{2}{*}{0.636 (1)} & U$_{11}$=0.0126 (12) \\
		& Al & 0.24 &  & U$_{33}$=0.0157 (12) \\
		\hline
		\multirow{2}{*}{2$a$} & Al & 0.76 & 0.410 (1) &U$_\mathrm{iso}$=0.0136 (23) \\
		& Pt & 0.24 & 0.376 (2) & U$_\mathrm{iso}$=0.0383 (65) \\
		\hline
		\multirow{3}{*}{4$b$} & \multirow{3}{*}{Al} & \multirow{3}{*}{1} & \multirow{3}{*}{0.251 (1)} & U$_{11}$=0.0173 (28) \\
		&  &  &  & U$_{22}$=0.0172 (28) \\
		&  &  &  & U$_{33}$=0.0151 (16) \\
		\hline\hline
	\end{tabular}
\end{table}

Cooling the crystal to 0.45\,K, the intensities and positions of selected reflections were monitored to verify whether CePtAl$_3$ undergoes any structural modifications. No significant changes were seen, suggesting that the peak observed in the specific heat at 3\,K is due to a magnetic transition that does not coincide with a strong response of the lattice. The intensities of the Bragg reflections also showed that the magnetic structure is characterized by a modulation vector $\vec{k} \neq 0$. The refinement of the 147 nuclear reflections measured at 0.45\,K strongly favors the disorder model established at ambient temperature. The refined parameters are close to those listed in \tab\ref{tab:crystal-structure}, \ie, the crystal structure is identical at 0.45\,K and ambient temperature, except for the displacement parameters, which at 0.45\,K are reduced by a factor of two, as expected for low temperatures.


\subsection{Spontaneous magnetic order}
\label{sect:magnetic-transition}

\begin{figure}[t]
	\begin{center}
		\includegraphics[width=0.48\textwidth]{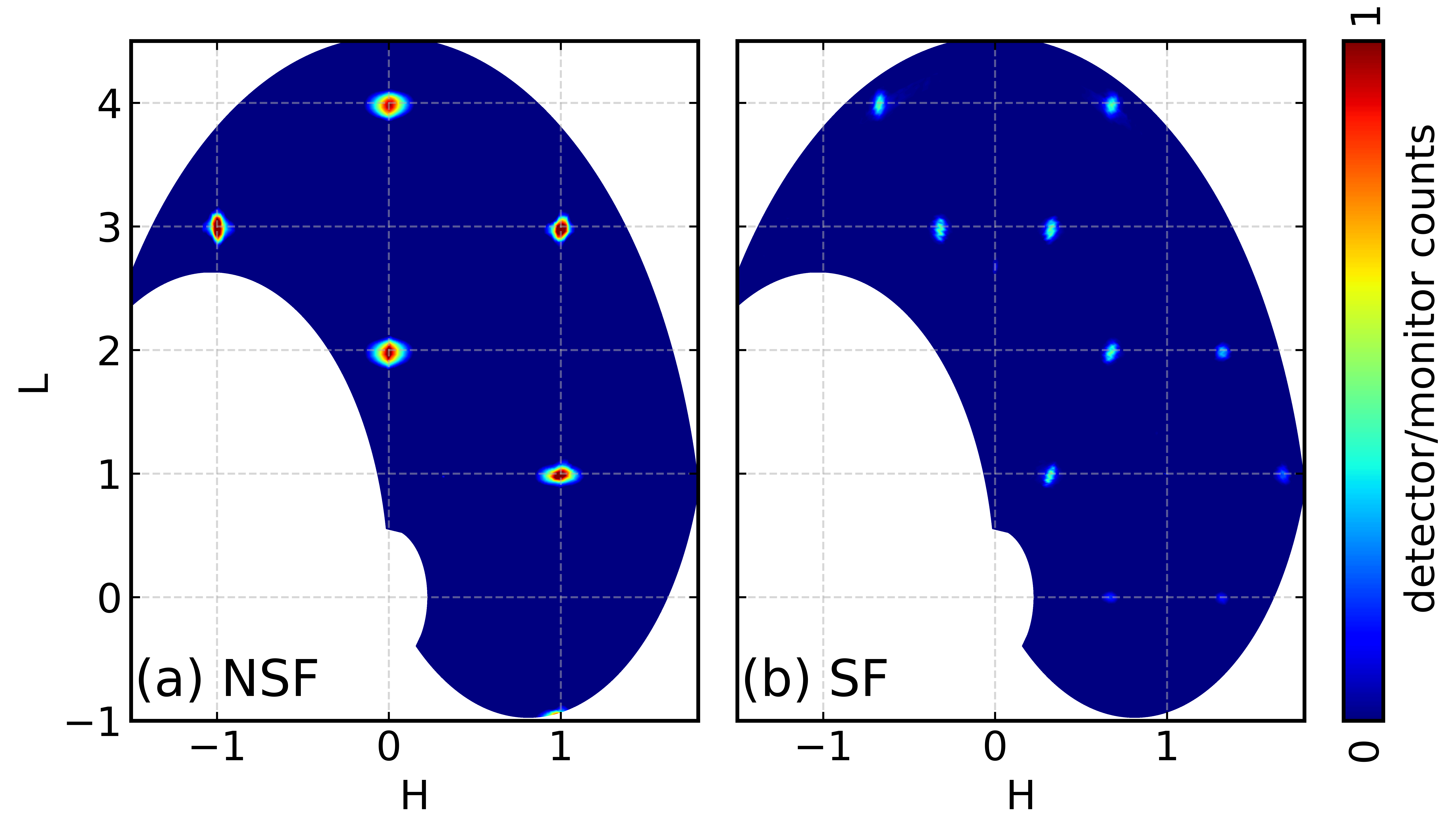}
	\end{center}
	\caption{Neutron diffraction intensity in the ($H0L$) plane of CePtAl$_3$ at 0.45\,K. (a) Non-spin-flip (NSF) channel, where nuclear reflections are observed. (b) Spin-flip (SF) channel, where magnetic reflections with a modulation vector $\vec{k}$=$(0.676\,0\,0)$ are observed.}
	\label{fig:dns}
\end{figure}

To determine the modulation vectors of the spontaneous magnetic order in CePtAl$_3$, \ie at zero magnetic field and low temperatures, we mapped the $(H0L)$ plane in reciprocal space using neutron diffraction at 0.45\,K. In addition to the nuclear reflections observed in the NSF channel, shown in \fig\ref{fig:dns}(a), 10 weak magnetic reflections were detected in the SF channel, as shown in \fig\ref{fig:dns}(b). The intensities of these magnetic reflections are roughly two orders of magnitude smaller than the nuclear reflections. They may be indexed by a modulation vector $\vec{k}$=$(0.676 (4)\,0\,0)$. The value $k_x$=0.676\,(4) is close to commensurate order with $k_x$=$\frac{2}{3}$=$0.\overline{6}$. However, further measurements confirmed that the modulation vector is, in fact, incommensurate.

\begin{figure}[t]
	\begin{center}
		\includegraphics[width=0.48\textwidth]{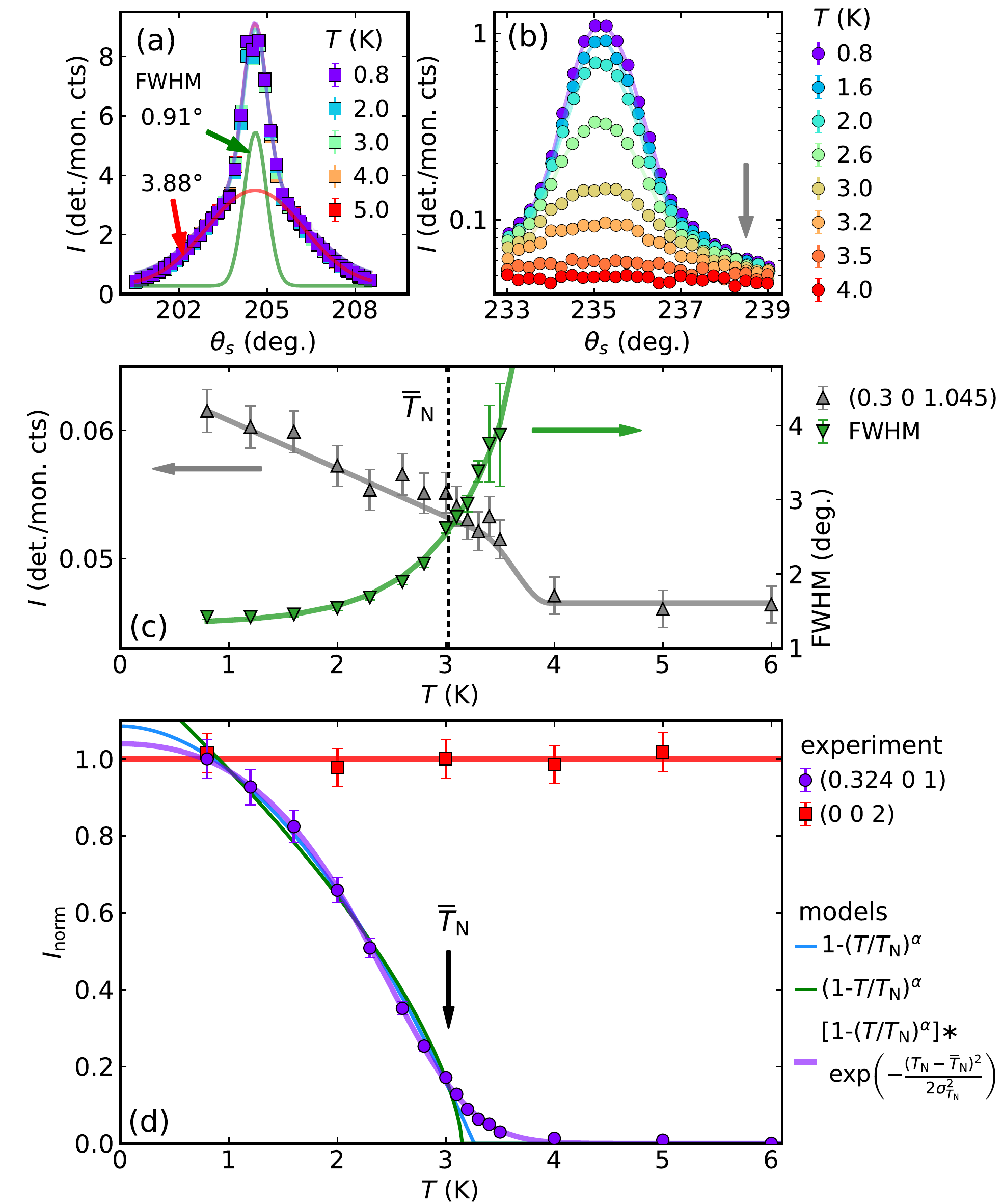}
	\end{center}
	\caption{Intensities of nuclear and magnetic reflections across the magnetic transition. (a) Nuclear $(0\,0\,2)$ reflection. It can be fitted with two Gaussian components (green and red lines) of different full width at half maximum (FWHMs), as indicated by arrows and labels. (b) Magnetic $(0.324\,0\,1)$ reflection. In (a) and (b), data are shown with points, and solid lines are guides to the eye. (c) Intensity in the vicinity of the magnetic reflection (gray triangles) at the position marked with an arrow in (b) and the FWHM of the magnetic reflection (green triangles) as a function of temperature. (d) Integrated intensity of reflections in (a) and (b) as a function of temperature. The intensity of the magnetic reflection is not accurately reproduced with a power-law behavior (blue and green lines). A power-law dependence convoluted with a Gaussian distribution (solid, purple line), that may be attributed to the structural inhomogeneities in the sample provides the best fit.}
	\label{fig:PANDA}
\end{figure}

The magnetic transition temperature was determined by means of neutron diffraction at PANDA. Upon cooling, the intensities of the magnetic and nuclear reflections at $(0.324\,0\,1)$ and $(0\,0\,2)$, respectively, were measured as shown in \fig\ref{fig:PANDA}. Below 3.5\,K, the intensity of the magnetic reflection increased, while the nuclear reflection did not change, indicating lack of an accompanying structural transition or sizable magnetostriction above the detection limit of our measurements, as mentioned in \sect\ref{sect:crystal}.

The nuclear reflection consisted of a sharp, resolution-limited diffraction peak with a FWHM of 0.91$^{\circ}$. The additional presence of diffuse shoulders is well described by a broad Gaussian with FWHM=3.88$^{\circ}$. We attribute the diffuse component to Huang scattering \cite{Huang-2000, Peisl-ActaCrystA-1975}. It arises from  the atomic disorder, which causes strain fields responsible for broad, diffuse signal contributions around some Bragg reflections. In comparison, the shape of the magnetic reflection is well described by a single Gaussian, without any qualitative change in line shape across the magnetic transition.

\begin{table}[t]
	\centering
	\caption{Parameters obtained from fitting the intensity of the $(0.324\,0\,1)$ magnetic reflection with different models shown in \fig\ref{fig:PANDA}(c).}
	\label{tab:models}
	\begin{tabular}{l   l l}
		\hline\hline
		Model & \multicolumn{2}{c}{Parameters} \\
		\hline
		$1-(T/T_\mathrm{N})^\alpha$ & $T_\mathrm{N}$=3.26 (3) K & $\alpha$=1.89 (11)  \\
		\hline
		$(1-T/T_\mathrm{N})^\alpha$ & $T_\mathrm{N}$=3.15 (6) K & $\alpha$=0.66 (5)   \\
		\hline		$[1-(T/T_\mathrm{N})^\alpha] \ast$ &$\overline{T}_\mathrm{N}$=3.02 (15) K & $\alpha$=2.7 (8) \\
		$\ast \mathrm{exp}(-(T_\mathrm{N}-\overline{T}_\mathrm{N})^2/2\sigma_{T_\mathrm{N}}^2)$ & $\sigma_{T_\mathrm{N}}$=0.48 (7) K & \\
		\hline\hline
	\end{tabular}
\end{table}

The temperature dependence of the intensity of the magnetic reflection may be equally well described by a power-law behavior $I \propto (1-T/T_N)^\alpha$ and a modified, heuristic expression $I \propto 1-(T/T_N)^\alpha$. Both functions are depicted in \fig\ref{fig:PANDA}(c) together with the measured intensities. Neither fit includes a point of inflection that accounts for the residual intensity above $T_\mathrm{N}$. Such residual intensity is commonly attributed to critical scattering that originates from slow, long-range correlations in the vicinity of the transition and gives rise to magnetic diffuse scattering. As a function of temperature, these correlations are most pronounced at the transition temperature \cite{Wilkinson-PhysRev-1956, Krivoglaz-1996}. In \fig\ref{fig:PANDA}(c) the width of the magnetic reflection and the intensity in the vicinity of the magnetic reflection as a function of temperature are shown. The width of the reflection increased with increasing temperature, consistent with critical fluctuations. However, the magnetic diffuse intensity measured at the position marked with an arrow in \fig\ref{fig:PANDA}(b), which corresponds to $(hkl)=(0.3\,0\,1.045)$, and shown in \fig\ref{fig:PANDA}(c) did not exhibit a maximum around $T_\mathrm{N}$. The diffuse scattering scales as the intensity of the magnetic reflection, suggesting static disorder of the long-range magnetic order, rather than fluctuations with a specific $Q$-dependence for the resolution of our setup. Considering the crystalline disorder characterized in \sect\ref{sect:crystal}, which reveals antisite disorder, the metallurgical appearance of our sample without hints for macroscopic phase segregation, and the broad peak in the specific heat, see \sect\ref{sect:cp}, we suggest therefore an alternative explanation of the residual scattering intensity above $T_\mathrm{N}$.

Namely, we propose that the atomic disorder and the structural inhomogeneities cause a spread of transition temperatures $T_\mathrm{N}$ that accounts for the residual scattering intensity above the transition, in addition to conventional critical scattering. The notion that structural inhomogeneities on the atomic scale may give rise to such a scattering contribution above the critical temperature was previously found to describe the magnetic transitions in polycrystalline samples of other materials \cite{Regulski-JPSJ-2004, ManchonGordon-Metals-2020}. Considering all information available, it represents perhaps the simplest scenario.

Incorporating the mathematical formalism reported by \citet{Regulski-JPSJ-2004}, we describe the spread of the transition temperatures seen in neutron scattering by a Gaussian distribution, with an average transition temperature $\overline{T}^n_\mathrm{N}$ and a standard deviation $\sigma^n_{T_N}$. The measured intensity is then a convolution of the power-law behavior with the Gaussian distribution as depicted in \fig\ref{fig:PANDA}(c). Using the modified power-law function $1-(T/T_\mathrm{N})^\alpha$, which proved to be insensitive to the temperature range fitted, the model is in excellent agreement with experiment across the entire temperature range, where $\overline{T}^n_\mathrm{N}$=3.03\,(15)\,K and $\sigma^n_{T_\mathrm{N}}$=0.48\,(7)\,K. It is worthwhile to point out that the convoluted model assumes only one additional parameter than a power-law, while significantly improving the quality of the fit.

The main conclusion suggested by this analysis also accounts for the broad signature in the specific heat shown in \fig\ref{fig:cp}, using an analogous model where an underlying specific heat curve is convoluted by the distribution of transition temperatures determined in neutron scattering. Although the exact parameterization of the specific heat is not straightforward, the width of the resulting convolution is dominated by the contribution from the distribution of transition temperatures \footnote{Convolution of two Gaussians with variance $\sigma^2$ and $\tau^2$, results in a Gaussian with variance $\sigma^2+\tau^2$. A broad contribution from a distribution of transition temperatures will hence dominate the convoluted Gaussian.}, \ie, the resulting curve depends little on the precise form of the parameterization in the disorder-free compound. Notably, the transition temperature $\overline{T}^{c_p}_\mathrm{N}$=3\,K and the FWHM of  $\approx$1.5\,K, which corresponds to a standard deviation $\sigma_{T_\mathrm{N}}^{c_p}$=0.64\,K observed in the specific heat, are very close to the values determined in neutron scattering, $\overline{T}^n_\mathrm{N}$=3.03\,(15)\,K and  $\sigma^n_{T_\mathrm{N}}$=0.48\,(7)\,K.


\subsection{Magnetic order under magnetic field}
\label{sect:magnetism}

\begin{figure}[t]
	\begin{center}
		\includegraphics[width=0.48\textwidth]{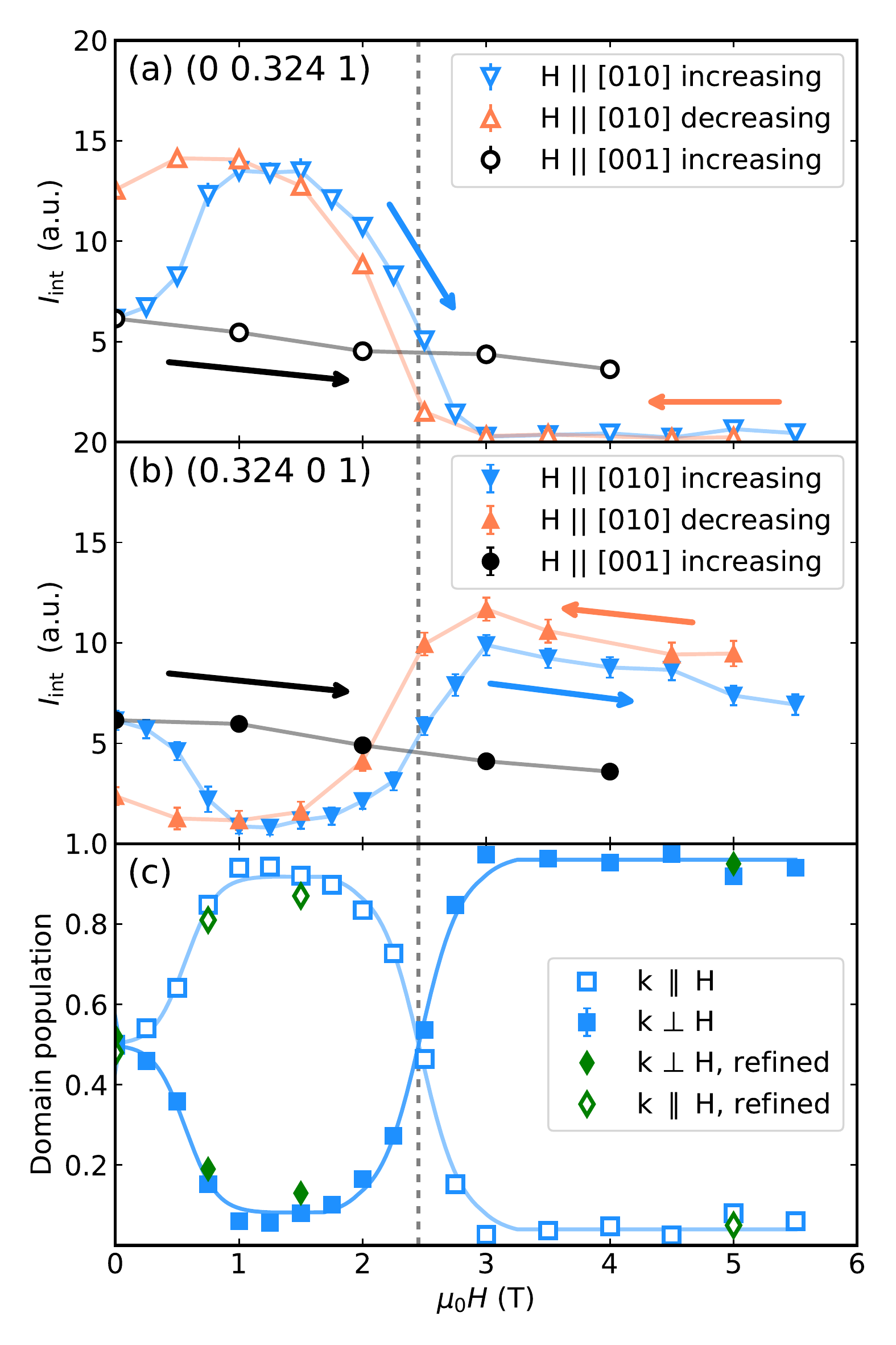}
	\end{center}
	\caption{Diffraction intensity as a function of magnetic field. Reflections were measured in increasing (blue) and decreasing (orange) field applied along the $[0\,1\,0]$ axis and in increasing field applied along the $[0\,0\,1]$ axis (black). (a) Intensity of the $(0\,0.32\,1)$ reflection (open symbols). (b) Intensity of the $(0.32\,0\,1)$ reflection (closed symbols). (c) Domain populations as determined by scaling the intensities of reflections in (a) and (b) under increasing field applied along the $[0\,1\,0]$ direction (blue squares) and determined from the refinement (green diamonds). Solid lines are guides to the eye. The vertical line marks the field of 2.5\,T, the mid point of domain repopulation.}
	\label{fig:magneticRefls}
\end{figure}

Various magnetic reflections were measured at 1.5~K as a function of magnetic field up to 5.5~T. The main observation inferred from these measurements is that the $(hkl)\pm(0.676 \, 0 \, 0)$ and $(hkl)\pm(0 \, 0.676 \, 0)$ families of magnetic reflections behave differently under applied magnetic field. The general trend may be illustrated by means of the $(0.324\,0\,1)$ and $(0\,0.324\,1)$ reflections shown in \fig\ref{fig:magneticRefls}.

Under increasing magnetic field applied along the $[0\,1\,0]$ direction (blue symbols in \fig\ref{fig:magneticRefls}), the intensity of the $(0\,0.324\,1)$ reflection increases and reaches a maximum with a plateau between 1 and 1.5\,T. Further increasing the field reduces the intensity of the $(0\,0.324\,1)$ reflection, which vanishes $>$3\,T up to 5.5\,T, the highest field studied. Subsequently decreasing the field to 1\,T (orange symbols in \fig\ref{fig:magneticRefls}), the intensity displays some hysteresis as highlighted by a curve guiding the eye. Below 1\,T, the intensity of the $(0\,0.324\,1)$ reflection does not drop to the value observed prior to increasing the field. Instead, it remains high, with a small drop just before the magnetic field reaches zero.

The behavior observed so far is contrasted by the intensity of the $(0.324\,0\,1)$ reflection, which is shown in \fig\ref{fig:magneticRefls}. Under increasing magnetic field applied along $[0\,1\,0]$ its intensity decreases almost to zero at 1\,T, followed by an increase, reaching a maximum at 3\,T, before slowly decreasing again when increasing the field further. Decreasing the field next, reveals the reverse evolution of the intensity. Analogous to the $(0\,0.324\,1)$ reflection, the intensity does not increase to reach the value before changing the field, but remains low with a small upturn just before the magnetic field reaches zero.

Turning now to the behavior for magnetic field along the $[0\,0\,1]$ direction (black symbols in \fig\ref{fig:magneticRefls}) the evolution of both reflections $(0.324\,0\,1)$ and $(0\,0.324\,1)$ is the same. Both exhibit essentially the same intensity and a monotonic decrease up to 4~T.

It is important to note that we did not observe any changes of the magnetic modulation vector as such or widths of the magnetic reflections for all applied fields studied under both increasing and decreasing field. The magnetic modulation vector determined under applied fields is $\vec{k}$=$(0.677\,(1) \; 0 \; 0)$, consistent with the value presented in \sect\ref{sect:magnetic-transition}. In addition, we did not observe significant intensity of the second-order mixed satellites, such as $(0.324\,0.676\,1)$ and others, that would be indicative of a multi-$k$ magnetic structure. Instead, the field dependence of the magnetic reflections is clearly characteristic of the formation of a multi domain, single-$k$ AFM structure. 

Namely, for a single-domain, multi-$k$ structure, the intensities of the $(0\,0.324\,1)$ and $(0.324\,0\,1)$ reflections would have to be the same in some field range, whereas a small field of order 0.25\,T already changes the intensities of these reflections, as shown in \fig\ref{fig:magneticRefls}. Further, up to 1\,T, the magnetic field changes the domain population, almost saturating the volume of the domain with the magnetic modulation vector along the field direction ($\vec{k} \parallel \vec{H}$), related to the intensity of the $(0\,0.324\,1)$ reflection.

Above 1.5\,T, an unexpected exchange of intensity may be observed between the two reflections and a re-population of the magnetic domains. Decreasing the field causes a hysteresis in the behavior of the intensities of the magnetic reflections. The difference of intensity after removing the applied magnetic field is, finally, characteristic of the presence of magnetic domains. Here, the relative intensities of the reflections allow us to determine the field dependence of the domain populations, as shown in \fig\ref{fig:magneticRefls}(c).


\subsection{Real-space depiction of magnetic order}

Establishing that CePtAl$_3$ exhibits a single-$k$ magnetic structure allows us to infer the orientation of the magnetic moments in real space. The single-$k$ character of the magnetic order lowers the symmetry of the tetragonal $I4mm$ space group to orthorhombic $Imm2$, where two irreducible representations $\sigma_1$ and $\sigma_2$ are allowed. For $\vec{k}$=$(0.676 \, 0 \, 0)$=$(k \, 0 \, 0)$, labeled $k_\perp$ since $\vec{k} \perp \vec{H} \parallel [0\,1\,0]$, the $\sigma_1$ representation describes an amplitude modulated antiferromagnet with the magnetic moment associated with the Ce atom at Wyckoff position $[x,y,z]$
\begin{equation}
	\label{eq:kx_sigma1}
	\vec{m}^{k_\perp}_{\sigma_1}(\vec{r}) = [0, m_1 \sin(2\pi k x), 0].
\end{equation}
In comparison, the $\sigma_2$ representation describes a cycloidal modulation of magnetic moments in the $ac$-plane, with a moment
\begin{equation}
	\label{eq:kx_sigma2}
	\vec{m}^{k_\perp}_{\sigma_2}(\vec{r}) = [m_2 \sin(2\pi k x), 0, m_3 \cos(2\pi k x)].
\end{equation}
The parameters $m_i$ describe the amplitude modulation along different axes. They may be refined using the experimental data. Note that, for the other domain with $\vec{k}$=$(0 \, 0.676 \, 0)$ labeled $k_\parallel$, the modulation vector direction is different, and the magnetic moments are given by 
\begin{equation}
	\label{eq:ky_sigma1}
	\vec{m}^{k_\parallel}_{\sigma_1}(\vec{r}) = [m_1 \sin(2\pi k y), 0, 0],
\end{equation}
and
\begin{equation}
	\label{eq:ky_sigma2}
	\vec{m}^{k_\parallel}_{\sigma_2}(\vec{r}) = [0, m_2 \sin(2\pi k y), m_3 \cos(2\pi k y)].
\end{equation}

\begin{table}[t]
	\centering
	\caption{Parameters of the CePtAl$_3$ magnetic structure determined from diffraction measurements at high magnetic field applied along the $[0\,1\,0]$ direction. Consecutive columns correspond to the magnetic field, domain population ratio, goodness of fit (GOF), $R$-factor for magnetic satellites, chosen representation, and magnetic moment components.}
	\label{tab:magnetic-structure}
	\begin{tabular}{l l l l l l}
		\hline\hline
		$\mu_0 H\,{\rm (T)}$ & $k_\perp$:$k_\parallel$  & GOF    &   $R_{sat}$    &   Repr.    & $m_i$ ($\mu_\mathrm{B}$)  \\ 
		\hline
		0 & 52:48     & 4.89    & 12.15  & $\sigma_2$ & $m_2$=0.852 (25) \\
		 & & & & & $m_3$=0.484 (37) \\ 
		\hline
		0.75 & 19:81     & 9.25    & 14.75  & $\sigma_2$ & $m_2$=0.836 (82)  \\ 
		 & & & & & $m_3$=0.494 (130) \\
		 \hline
		1.5 & 13:87  & 8.44   & 16.71 & $\sigma_2$ & $m_2$=0.846 (77)  \\ 
		 & & & & & $m_3$=0.52 (12) \\
		 \hline
		5  & 95:5   & 13.5   & 8.99  & $\sigma_1$ & $m_1$=0.64 (15)   \\
		\hline
		5  & 95:5   & 13.62  & 8.12  & $\sigma_2$ & $m_2$=0.65 (16) \\ 
		 & & & & &  $m_3$=0 (0.5) \\
		 \hline\hline
	\end{tabular}
\end{table}

\begin{figure}[t]
	\begin{center}
		\includegraphics[width=0.48\textwidth]{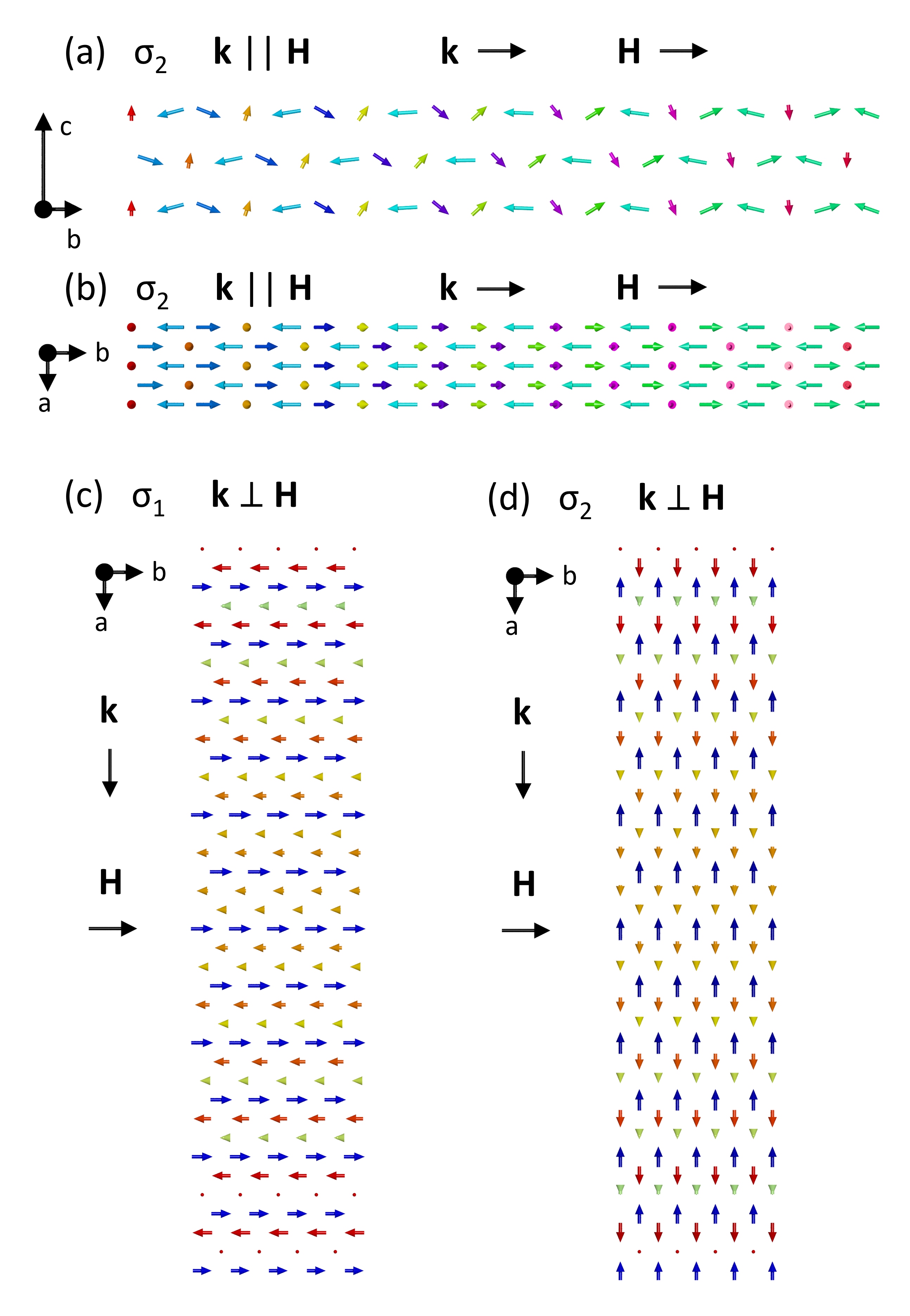}
	\end{center}
	\caption{Depiction of the incommensurate component of the magnetic order of CePtAl$_3$ under applied magnetic field. (a) and (b) Cycloidal phase with $\vec{k} \parallel \vec{H}$ observed $\le$2.5\,T, viewed along (a) the $[1\,0\,0]$ direction and (b) the $[0\,0\,1]$ direction. (c) and (d) Amplitude-modulated phase with $\vec{k} \perp \vec{H}$ observed between 2.5 and 5.5\,T with the moments oriented along (c) the $[0\,1\,0]$ direction or (d) the $[1\,0\,0]$ direction; view along the $[0\,0\,1]$ direction ($c$ axis). On the left-hand side of each panel, a coordinate system is shown which corresponds to the size of a nuclear unit cell. Next to each panel, the directions of the modulation vector $\vec{k}$ and the applied magnetic field $\vec{H}$ are shown. For the cycloidal structure, colors mark the azimuthal angle between the magnetic moment and the $[0\,0\,1]$ direction. For the amplitude-modulated structure, colors mark the modulus of the magnetic moment.}
	\label{fig:magneticOrder}
\end{figure}

The refinement of the dataset collected at $\mu_0 H$=0, 0.75 and 1.5\,T strongly supports a cycloidal structure, where the refined magnetic moment components are listed in \tab\ref{tab:magnetic-structure}, and the arrangement of the moments in real space is shown in \fig\ref{fig:magneticOrder}.
For the data collected at 5\,T the refinement is not conclusive, and both cycloidal and amplitude-modulated magnetic structures account for the data equally well.

One of the causes of this uncertainty may be the relatively small number of reflections measured.
For the cycloidal model, the $m_3$ parameter, which is the modulation amplitude along the $c$ axis, is poorly constrained, and the best refinement is obtained when setting it to zero.
As a result, at 5\,T, both models correspond to the amplitude-modulated magnetic structure with moments lying either along the $b$ or $a$ crystal axis, for the $\sigma_1$ and $\sigma_2$ models, respectively, as shown in \fig\ref{fig:magneticOrder}.
Since high magnetic fields tend to align the microscopic magnetic moments along the field direction, the model with the $\sigma_1$ representation seems more likely.
However, further investigations will be necessary to resolve these uncertainties.


\subsection{Further discussion}

The evolution of the magnetic structure in CePtAl$_3$ under magnetic field applied along the $[0\,1\,0]$ direction may be summarized as follows. At $\mu_0 H$=0, two equally populated magnetic domains coexist, supporting cycloidal magnetic order described by \eq{\ref{eq:kx_sigma2}} and \eq{\ref{eq:ky_sigma2}}. One domain exhibits a modulation vector $\vec{k_\perp} = (0.676 \, 0 \, 0)$, while the other exhibits $\vec{k_\parallel} = (0 \, 0.676 \, 0)$. 

Application of the magnetic field along the $[0\,1\,0]$ direction changes the domain populations, initially favoring the domain with $\vec{k_\parallel}$=$(0 \, 0.676 \, 0)$. This domain population becomes saturated between 1 and 1.5\,T, where the magnetic moments are described by \eq{\ref{eq:ky_sigma2}} as depicted in \fig\ref{fig:magneticOrder}(a) and \fig\ref{fig:magneticOrder}(b). Further, between 1.5 and 3\,T, an exchange of the intensity is observed between the magnetic reflections with $\vec{k_\parallel}$ and $\vec{k_\perp}$, as shown in \fig\ref{fig:magneticRefls}(c), corresponding to a domain repopulation. This repopulation of the domains is associated with a transition to an amplitude modulated structure, where the moments are either oriented along $[1\,0\,0]$ or $[0\,1\,0]$ with a modulation amplitude $m\approx$\,0.65\,$\mu_\mathrm{B}$. In contrast, when the magnetic field was applied along $[0\,0\,1]$, we did not see any evidence suggesting a reorientation of the magnetic moments, and the domains remained equally populated at all applied magnetic fields.

For magnetic field exceeding 3\,T, the intensities of the magnetic reflections decreased monotonically with increasing field for both field directions. This is consistent with a reduction of the modulation amplitude and the increase of the uniform magnetization detected in the magnetization \cite{Franz-PhD}. Indeed, we observed an increase of intensity at specific $\vec{k}=0$ reflections, reflecting the uniform magnetization in the scattering process. Moreover, for field applied along the $[0\,1\,0]$ direction, the magnetization measured at 2\,K and 5\,T of 0.75\,$\mu_\mathrm{B}$ per cerium atom \cite{Franz-PhD} was of the same order of magnitude as the amplitude of the modulation, $m_3$=0.65\,$\mu_\mathrm{B}$, inferred from our neutron diffraction measurements. Overall, the magnetic structure is consistent with the magnetic anisotropies determined from previous measurements \cite{Franz-JAC-2016, Franz-PhD}, which establish an easy magnetization plane $(0\,0\,1)$ and a hard magnetic axis along the $[0\,0\,1]$ direction, since the main components of the modulated part of the magnetization lie in the $(0\,0\,1)$ plane, as summarized in \tab\ref{tab:magnetic-structure}.

Most likely, the tilting of the magnetic moment away from the easy $(0\,0\,1)$ plane is the result of Dzyaloshinskii–Moriya interactions allowed for the non-centrosymmetric $I4mm$ space group, see Tab. II in Ref.\,\onlinecite{Bauer-PRM-2022}. Alternatively, the field dependence may suggest coexisting local and itinerant magnetism in CePtAl$_3$, as reported for other cerium compounds \cite{Nakano-PRB-2019, Hoshino-PRL-2012}, where the itinerant electrons might carry the magnetic moment component along the $c$ axis, driving the incommensurability of the magnetic order. 

Within the experimental resolution of our study, we did not observe any changes in the length scales of the magnetic modulations reflected in the constant component of the modulation vector $k$=$0.677\,(1)$ under applied field. Our experimental setup allowed us to clearly distinguish that the modulation is not commensurate with $k$=$\frac{2}{3}$. The mechanism causing the formation of the incommensurability and the pinning of the magnetic modulation length cannot be explored further in our study. Since it is unchanged for both field directions and up to relatively high magnetic fields, it seems possible that the pinning involves structural disorder and follows the underlying atomic length scales of the disorder.


\section{Conclusions}

In conclusion, we reported on a detailed neutron diffraction study of the crystal and magnetic structure of CePtAl$_3$ complemented by specific heat measurements. For our study, we have grown a large single crystal of CePtAl$_3$ by means of optical floating-zone method, which exhibits occupational and positional disorder on atomic length scales in neutron diffraction measurements. In the simplest scenario, the disorder affects the long-range magnetic order, which is described well in terms of a spread of the transition temperatures with an average transition temperature $\overline{T}^n_\mathrm{N}$=3.03\,(15)\,K and standard deviation $\sigma^n_{T_\mathrm{N}}$=0.48\,(7)\,K. This model is quantitatively consistent between the temperature dependence of the magnetic order parameter observed in neutron diffraction and the distribution of anomalies in the specific heat at the magnetic transition where $\overline{T}^{c_p}_\mathrm{N}$=3\,K and $\sigma^{c_p}_{T_\mathrm{N}}$=0.64\,K. 

Measurements under applied magnetic field provided further information on the magnetic structure of CePtAl$_3$. For fields up to 1\,T applied along $[0\,1\,0]$, CePtAl$_3$ exhibits the characteristics of a multi domain state with single-$k$, incommensurate, cycloidal magnetic order. Between 1 and 1.5\,T, a single domain state is stable. Further increasing the field causes a magnetic transition between 1.5 and 3\,T, suggesting a change from cycloidal to amplitude-modulated magnetic order. Above 3\,T, the magnetic order may be described by amplitude-modulated moments in the tetragonal plane. For fields applied along $[0\,0\,1]$, no transitions were observed, and CePtAl$_3$ remains in a multi domain state up to 4\,T, the highest field studied. Putative mechanisms at the heart of the AFM order may be Dzyaloshinskii–Moriya spin-orbit interactions or, more speculatively, a mixed itinerant and local-moment magnetism carried by the conduction and 4$f$-electrons, respectively.

Taken together, the evidence for structural disorder and smearing of the magnetic phase transition in CePtAl$_3$ at zero magnetic field and the magnetic field dependence of the AFM structure from cycloidal to amplitude modulated order identify a highly unusual interplay of magnetic order, magnetic textures, and structural properties in the presence of disorder in an $f$-electron compound.


\section{Acknowledgements}

We kindly acknowledge Thomas M\"{u}ller for help with the analysis of the DNS data, Alistair Cameron for support at D23, as well as Marc Wilde and Vivek Kumar for helpful discussions. We also thank the staff at the MLZ and ILL for support. Parts of the neutron measurements were performed at POLI, jointly operated by the RWTH Aachen University and JCNS within the JARA-FIT collaboration. 
 
This study was funded by the Deutsche Forschungsgemeinschaft under Project No. WI33203-1, TRR80 (From Electronic Correlations to Functionality, Project No.\ 107745057, Project E1), SPP2137 (Skyrmionics, Project No.\ 403191981, Grant No. PF393/19), and the excellence cluster MCQST under Germany's Excellence Strategy EXC-2111 (Project No.\ 390814868). Financial support by the European Research Council through Advanced Grants No.\ 291079 (TOPFIT) and No.\ 788031 (ExQuiSid) and through the European Union's Horizon 2020 research and innovation program as well as the Czech Science Foundation GA\v{C}R under the Junior Star grant No. 21-24965M (MaMBA) is gratefully acknowledged. 

%


\end{document}